\documentclass[prl,superscriptaddress,twocolumn]{revtex4}

\usepackage{amsfonts,amssymb,amsmath}
\usepackage[]{graphics,graphicx,epsfig}
\usepackage{amsthm}

\def\identity{\leavevmode\hbox{\small1\kern-3.8pt\normalsize1}}

\renewcommand{\epsilon}{\varepsilon}

\begin{document}

\title{Generalized monogamy of contextual inequalities from the no-disturbance principle}
\author{Ravishankar \surname{Ramanathan}}
\affiliation{Centre for Quantum Technologies, National University of Singapore, 3 Science Drive 2, 117543 Singapore, Singapore} 
\author{Akihito \surname{Soeda}}
\affiliation{Centre for Quantum Technologies, National University of Singapore, 3 Science Drive 2, 117543 Singapore, Singapore}
\author{Pawe{\l} \surname{Kurzy\'nski}}
\affiliation{Centre for Quantum Technologies, National University of Singapore, 3 Science Drive 2, 117543 Singapore, Singapore}
\affiliation{Faculty of Physics, Adam Mickiewicz University, Umultowska 85, 61-614 Pozna\'{n}, Poland} 
\author{Dagomir \surname{Kaszlikowski}}
\email{phykd@nus.edu.sg}
\affiliation{Centre for Quantum Technologies, National University of Singapore, 3 Science Drive 2, 117543 Singapore, Singapore}
\affiliation{Department of Physics, National University of Singapore, 2 Science Drive 3, 117542 Singapore, Singapore}

\begin{abstract}
In this paper we demonstrate that the property of monogamy of Bell violations seen for no-signaling correlations in composite systems can be generalized to the monogamy of contextuality in single systems obeying the Gleason property of no-disturbance. We show how one can construct monogamies for contextual inequalities by using the graph-theoretic technique of vertex decomposition of a graph representing a set of measurements into subgraphs of suitable independence numbers that themselves admit a joint probability distribution. After establishing that all the subgraphs that are chordal graphs admit a joint probability distribution, we formulate a precise graph-theoretic condition that gives rise to the monogamy of contextuality. We also show how such monogamies arise within quantum theory for a single four-dimensional system and interpret violation of these relations in terms of a violation of causality. These monogamies can be tested with current experimental techniques.
\end{abstract}
\maketitle

{\it Introduction.} The essence of the classical description of Nature is the assumption that the physical world exists independently of any observers and that the act of observation does not disturb it. This feature, called {\it realism} was first brought to mainstream physics by Einstein, Podolsky and Rosen (EPR) in \cite{EPR}. A mathematical consequence of realism is that there exists a joint probability distribution for the outcomes of measurements for all physical properties of the system \cite{Fine}. It was shown by Bell \cite{Bell} and by Kochen and Specker \cite{KS} that the property of realism is not present in quantum theory. Therefore, it is reasonable to treat realism as a hallmark of classicality and the lack of it as an indicator of quantumness. Clauser {\it et al.} \cite{CHSH} and Klyachko {\it et al.} \cite{Klyachko} proposed minimal experimental tests of this feature by means of the Bell-CHSH inequality and the Klyachko-Can-Binicoglu-Shumovsky (KCBS) inequality respectively. These inequalities must be satisfied in any theory that incorporates realism and their violations have been observed experimentally \cite{Aspect, Zeilinger}, confirming that Nature is not compatible with realism. 

 
Although Bell inequalities are studied as indicators of ``local realism" and KCBS inequalities are treated as indicators of ``non-contextuality", they have the same root, namely the assumption of realism. Local realism is a special kind of realism where the additional constraint of locality is imposed, namely that measurements in spatially separated systems do not influence each other. Bell inequalities thus require at least two correlated and spatially separated subsystems. The notion of non-contextuality on the other hand applies to a single system, and stipulates that the outcomes of any measurement are independent of any other measurement that can be jointly performed with it. Violation of the KCBS inequality implies that the system does not admit a non-contextual description for these measurements. Contextual inequalities being applicable to single systems are arguably more fundamental than Bell inequalities in studying the quantumness of physical systems. 

Quantum correlations as captured by the violation of Bell inequalities have been shown to obey the interesting property of monogamy \cite{Pawlowski}; if Alice is able to violate a Bell inequality with Bob, she is unable to violate the same Bell inequality with Charlie. This property only arises under certain conditions though, namely: i) Alice uses the same settings to violate Bell inequalities with both Bob and Charlie; ii) No communication between Alice, Bob and Charlie is allowed; iii) Bob and Charlie cannot use more measurement settings than two; iv) Alice tries to violate the very same Bell inequality with both Bob and Charlie. Bell monogamies are useful in secure quantum key distribution \cite{Pawlowski2}, interactive proof systems \cite{Toner} and in the emergence of a local realistic description for correlations in the macroscopic domain \cite{MacroBell}.

The fact that the origin of Bell inequalities and contextual inequalities is the existence of joint probability distributions suggests that a similar monogamy relation may hold for contextual inequalities as well. Bell monogamy arises as a consequence of the principle of no-signaling, which states that the probabilities of outcomes of measurement in one subsystem are independent of the choice of measurement in a spatially separated subsystem. An interesting question is how the properties of no-signaling and monogamy translate to contextual inequalities.

In this Letter, we focus on KCBS-type contextual inequalities and show that there is a form of monogamy of their violations analogous to the monogamy of Bell inequality violations. We exploit the Gleason principle of no-disturbance that is a generalization of the principle of no-signaling. The paper is organized as follows. We first formulate mathematically the principle of no-disturbance. We then derive a monogamy relation for the simplest contextual inequality, namely the KCBS inequality for five measurements and show that it applies to any theory that obeys the principle of no-disturbance, in particular quantum mechanics. Using techniques of graph theory, we show how these monogamies can be generalized to any contextual and Bell inequalities. In particular, we establish a Proposition that any chordal graph representing a set of measurements admits a joint probability distribution and use this to identify the necessary and sufficient condition for a set of measurements to exhibit the monogamy of contextuality.

{\it Principle of no-disturbance.} To formulate the principle of no-disturbance mathematically, let us consider a physical system on which one can perform several different measurements $A, B, C,$ etc. Let us assume that measurements $A$ and $B$ can be jointly performed as can measurements $A$ and $C$. This implies the existence of the joint probabilities $p(A = a,B = b)$ and $p(A = a,C = c)$ (where small letters denote outcomes of the corresponding measurements). The principle of no-disturbance is then the condition that the marginal probability $p(A = a)$ calculated from $p(A=a,B=b)$ is the same as that calculated from $p(A=a,C=c)$,  i.e., 
\begin{equation}
\sum_{b}p(A=a,B=b)=\sum_{c}p(A=a,C=c)=p(A=a).
\label{ND} 
\end{equation}
This property has been referred to as the Gleason property in \cite{Cabello} since it is the condition underlying Gleason's theorem. Note that when measurements $B, C$ are performed on spatially separated systems the principle of no-disturbance reduces to that of no-signaling. From here on we use $p(A=a,B=b)$ and $p(a,b)$ interchangeably wherever there is no possibility of confusion. 

{\it Monogamy of KCBS-type inequalities.} We concentrate first on the  KCBS inequality from \cite{Klyachko} that was introduced to test the quantumness of a single (three-level) system and construct a monogamy relation for it. Similar monogamies hold for any inequalities of this kind \cite{Entropic}. The KCBS inequality reads
\begin{equation}\label{e1}
\sum_{i=1}^{5} p(A_i=1) \leq 2,
\end{equation}
where $A_i$ are measurements with outcomes $a_i = 0,1$. These measurements are cyclically compatible (i.e., it is possible to experimentally determine $p(a_i, a_{i+1})$ (where one identifies $a_6$ with $a_1$)) and exclusive (i.e., $a_i a_{i+1}=0$). Measurements $A_{i-1}$ and $A_{i+1}$ are said to provide two different contexts for the measurement $A_i$.
These measurements can be represented by the ``commutation graph" corresponding to a pentagon where the vertices of the pentagon graph represent the five measurements and edges between any two vertices indicate that the two corresponding measurements can be jointly performed and are mutually exclusive. The bound of $2$ is derived under the assumption of existence of the joint probability distribution $p(a_1, a_2, \dots, a_5)$. In graph theoretic terms this bound corresponds to the independence number of the pentagon graph, which is the maximum number of mutually disconnected vertices in the graph. This inequality is violated in any contextual theory such as quantum theory where such a joint probability distribution does not exist. The KCBS inequality (\ref{e1}) is the necessary and sufficient condition for the existence of non-contextual description for these five measurements. Analogous inequalities can be constructed for larger number of measurements as well \cite{Cabello, Entropic}.

\begin{figure}[t]
\vspace{-1cm}
\begin{center}
\includegraphics[scale=0.2]{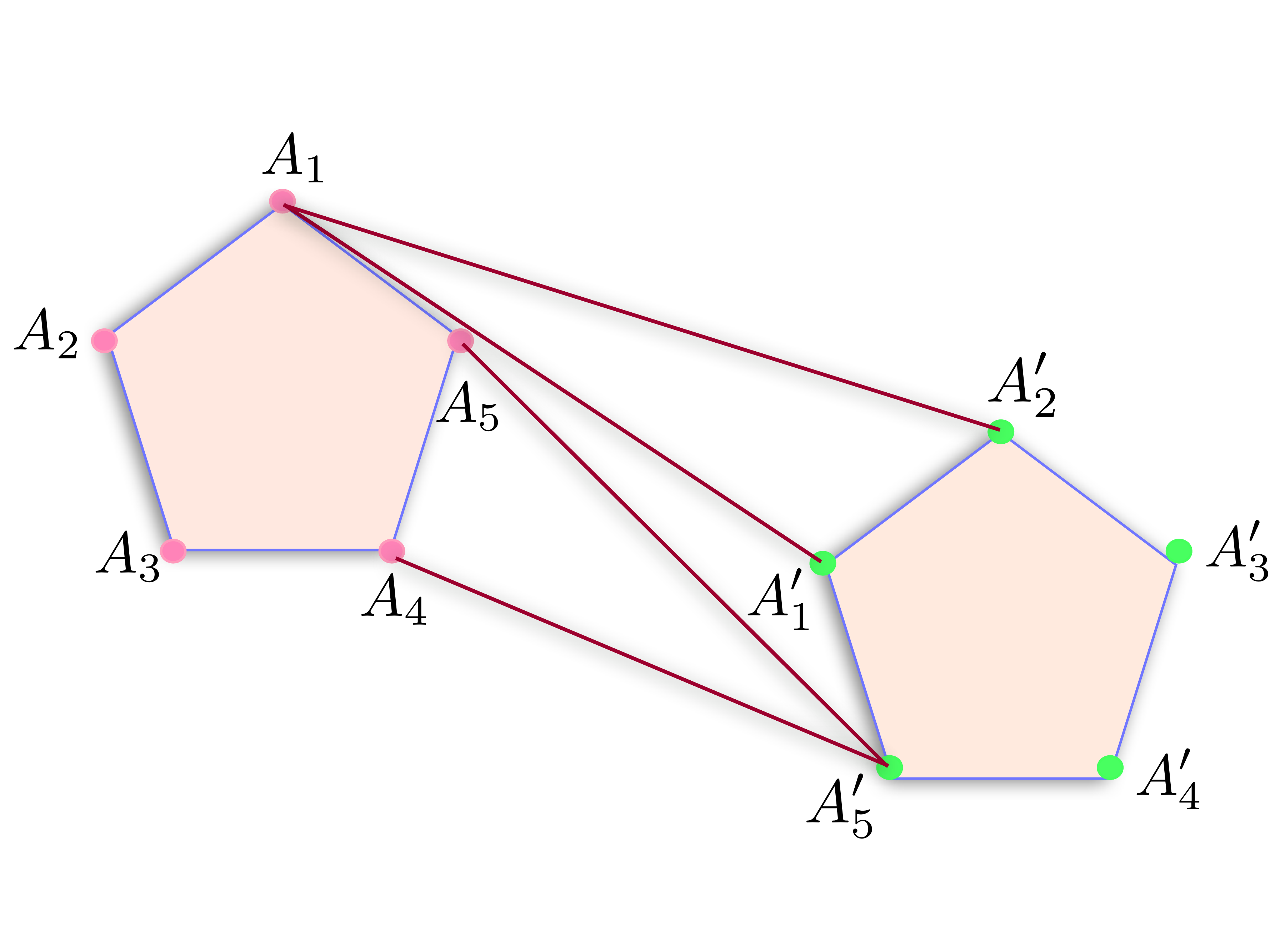}
\end{center}
\vspace{-1cm}
\caption{Graphical representation of two KCBS inequalities that satisfies the monogamy relation.}
\label{figvect}
\end{figure}

We can now precisely state the definition of contextual monogamy as the following. A set of measurements is said to have `monogamous contextuality' if it can be partitioned into disjoint subsets, each of which can by themselves reveal contextuality, but which cannot all simultaneously be contextual. Let us derive a monogamy relation for the KCBS contextual inequality from the no-disturbance principle, along similar lines to the monogamy of Bell inequality violations derived from the no-signaling principle \cite{Pawlowski}. Consider two sets of cyclically compatible and exclusive measurements $\{A_i\}$ and $\{A'_i\}$. Each set gives rise to a KCBS inequality (\ref{e1}). Let us assume that the triple $A_1, A'_1, A'_2$ are jointly measurable and mutually exclusive, as is also the triple $A_4, A_5, A'_5$. This scenario is represented by the commutation graph in Fig. \ref{figvect}. Therefore, in addition to $p(a_i,a_{i+1})$ and $p(a'_i, a'_{i+1})$, one can experimentally determine probabilities $p(a_1,a'_1,a'_2)$ and $p(a'_5,a_4,a_5)$. This condition is similar to a condition imposed in the derivation of Bell monogamies, namely that a common observer chooses the same settings for the violation of Bell inequalities with all other observers. 

We introduce the no-disturbance principle (\ref{ND}) by setting $p(A_1=1) = p$ and $p(A'_5=1) = q$. Mutual exclusiveness implies that $p(A'_1=1)+ p(A'_2=1) \leq 1-p$ and $p(A_4=1) + p(A_5=1) \leq 1-q$ in addition to $p(A_i=1) + p(A_{i+1}=1) \leq 1$ and $p(A'_{i}=1) + p(A'_{i+1}=1) \leq 1$. However, this already implies $
\sum_{i=1}^{5} p(A_i=1) \leq 2-q+p$ and $\sum_{i=1}^{5} p(A'_i=1) \leq 2-p+q$
and therefore the monogamy relation 
\begin{equation}
\sum_{i=1}^5 p(A_i=1) + \sum_{i=1}^5 p(A'_i = 1) \leq 4
\label{e2}
\end{equation} 
holds. Therefore, only one KCBS inequality out of the two sets $\{A_i\}$ and $\{A'_i\}$ can be violated in all theories that obey the no-disturbance principle such as quantum mechanics. 

If however the principle of no-disturbance does not hold, it is possible to violate both inequalities simultaneously. This is because in this case $p(a_1)$ calculated from $p(a_1,a'_1,a'_2)$ would yield a different value than that calculated from $p(a_1,a_5)$ or $p(a_1,a_2)$ (similarly for $A'_5$). The consequence of this would be that causality is violated, as can be seen from the following argument. In order to evaluate probabilities, joint measurements do not have to be performed simultaneously, they can as well be performed in sequential order. The fact that $p(a_1)$ depends on whether $A_1$ was measured with $A'_1, A'_2$ or with $A_2$ or $A_5$ can be used to signal backward in time. The marginal probabilities $p(a_1)$ calculated from $p(a_1,a'_1,a'_2)$ being not consistent with the probability $p(a_1)$ measured earlier (in a joint measurement of $A_1$ and $A_2$ for instance) would imply an influence propagating backward in time, thus violating causality \cite{Peres}. The no-signaling principle being a special instance of the no-disturbance principle, violation of no-signaling monogamies for Bell inequalities would imply the possibility of superluminal communication between spatially separated systems, which could also lead to a violation of causality. In the Supplementary Material, we help clarify the analogous relationship between the monogamy of contextual inequalities derived here and the monogamy of Bell inequalities (which are special instances of contextual inequalities) by deriving the monogamy of Bell-CHSH inequalities using the commutation graph technique introduced above.

Having illustrated the method for deriving monogamy relations for contextual (and Bell) inequalities we now proceed to formulate it using some graph-theoretic notions. To do so, we first state the following Proposition $1$ whose proof is provided in the Supplementary Material. 

{\it Proposition $1$:} A commutation graph $G$ representing a set of $n$ measurements (for any $n$) admits a joint probability distribution for these measurements if it is a chordal graph.

A chordal graph is a graph that does not contain an induced cycle of length greater than $3$, i.e., each of the cycles of four or more vertices in the graph must have a chord, an edge connecting two non-adjacent vertices in the cycle. This class of graphs comprises a large class of all graphs of $n$ vertices and the above Proposition $1$ excludes the construction of contextual inequalities (or Kochen-Specker proofs) from all such graphs. 

We can now precisely formulate the method for the derivation of monogamy relations introduced previously. Given a commutation graph representing a set of $n$ contextual inequalities, look for its vertex decomposition into $m$ chordal subgraphs (each of which admits a joint probability distribution by the Proposition $1$), such that the sum of the independence numbers of these subgraphs is $n*R$, where $R$ is the non-contextual bound for each of the inequalities. If the $n$ contextual inequalities are not all the same, i.e., if $n_1$ inequalities have non-contextual bound $R_1$, $n_2$ inequalities have bound $R_2$ etc., then the subgraphs should be chosen such that the sum of their independence numbers is $\sum_k n_k R_k$. All vertices of the commutation graph are to be included in the vertex decomposition into subgraphs with no vertex appearing in more than one subgraph, but the edges between the different subgraphs can be neglected. Note that while many contextual inequalities involve rank-1 projectors, where the edges of the graph denote mutual exclusiveness in addition to compatibility, this assumption is not crucial to the derivation of monogamies. This can be seen from the derivation of the Bell inequality monogamies, where only compatibility is required.

\begin{figure}[t]
\begin{center}
\includegraphics[scale=0.32]{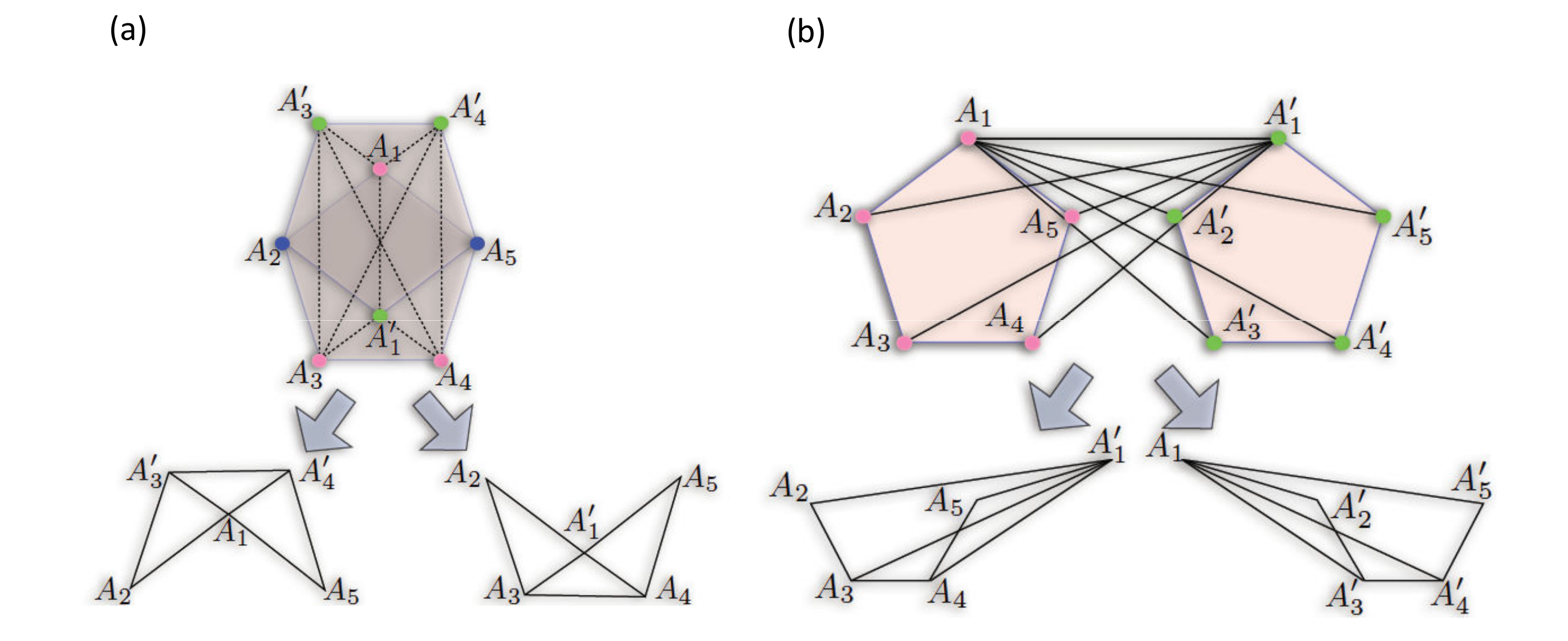}
\end{center}
\vspace{-0.6cm}
\caption{Measurement configurations (top) and their decompositions (bottom) for which monogamy relations can be derived.}
\label{fig3}
\end{figure}

Using the method presented above, one can identify several commutation graphs that yield contextual monogamy (including Bell monogamy) relations, for instance the monogamy relation (\ref{e2}) also holds for the graphs in Fig. \ref{fig3}(a) and Fig. \ref{fig3}(b) as can be seen by the decompositions given there. We see that monogamy relations for two KCBS inequalities can be derived for various measurement configurations, the measurement configuration given in Fig. \ref{figvect} being the minimal one (with fewest edges connecting two contextual graphs) in which such monogamies appear for two sets of five separate measurements. This minimality can be seen by finding that for all graphs with one, two and three edges connecting two distinct pentagon graphs, no vertex decomposition into two or more chordal subgraphs with total independence number $4$ exists. Since the KCBS inequality (\ref{e1}) is a necessary and sufficient condition for the existence of non-contextual description for the five measurements, the relation (\ref{e2}) holds for any contextual inequality of this kind as well \cite{Entropic}. The method can also be used to construct monogamy relations for inequalities with more than five measurements along similar lines. In general, it can be seen that the larger the number of mutually exclusive and jointly performable measurements, the stronger the monogamy relation is. 

We now proceed to explicitly identify the commutation graphs that give rise to monogamy relations for a given set of $n$ KCBS-type contextual inequalities (with non-contextual bound $R$). This is done in the Proposition $2$ which provides the necessary and sufficient condition for a commutation graph to give rise to a monogamy relation using the method outlined above, its proof is provided in the Supplementary Material.

{\it Proposition 2:} Consider a commutation graph representing a set of $n$ KCBS-type contextual inequalities each of which has non-contextual bound $R$. Then this graph gives rise to a monogamy relation using the outlined method if and only if its vertex clique cover number is $n*R$. 

The vertex clique cover number is the minimal number of cliques required to cover all the vertices of the graph. The above Proposition can be extended to the case when when one is interested in the monogamy of a set of $n_k$ different contextual inequalities with different non-contextual bounds $R_k$, with $\sum_k n_k = n$. Then the condition becomes that the vertex clique cover number equal $\sum_k n_k R_k$. This gives a very powerful method of identifying whether a given graph exhibits contextual monogamy. 

The monogamy relations presented so far are genuine properties of contextual theories in the sense that for classical (non-contextual) theories each of the two inequalities can achieve its maximal value.  We note, however, that certain monogamies also hold for non-contextual theories, for instance for the situation when all measurements in set $\{A_i\}$ are compatible and mutually exclusive with all the measurements in set $\{A'_i\}$. Here the mutual exclusiveness guarantees the monogamy $\sum_{i=1}^{5} p(A_i = 1) + \sum_{i=1}^{5} p(A'_i = 1) \leq 5/2$ in all theories obeying the no-disturbance principle. However, an important feature here is that monogamies also arise within non-contextual theories for which the relation $\sum_{i=1}^{5} p(A_i = 1) + \sum_{i=1}^{5} p(A'_i = 1) \leq 2$ holds, and this can be traced to the large number of mutually exclusive measurements required here. The interesting monogamies are those in which such classical restrictions do not appear such as those in Fig. \ref{figvect} and Fig. \ref{fig3}.


Let us now show how the monogamy relation (\ref{e2}) applies within quantum theory. Firstly, note that measurements for the optimal violation of KCBS inequality for a single three-level quantum system are $rank-1$ projectors spanning real three-dimensional space. Consider a real four-dimensional space in which the set of projectors $\{A_i\}$ spans dimensions $1,2$ and $3$ and the set of projectors $\{A'_{i}\}$ spans dimensions $2, 3$ and $4$. These projectors can be constructed to obey the constraints of mutual exclusiveness and joint measurability as required by the commutation graphs. A set of projectors that correspond to the measurement configuration in Fig. 3(b) for a quantum mechanical system of dimension four is given by
\begin{eqnarray}
& &|A_1\rangle = (1,0,0,0)^T,~|A_2\rangle = (0,1,0,0)^T, \nonumber \\
& &|A_3\rangle = (\cos\theta,0,\sin\theta,0)^T, \nonumber \\ & &|A_4\rangle = (\sin\alpha\sin\theta, \cos\alpha, -\sin\alpha \cos\theta, 0 )^T  \nonumber \\
& &
|A_5\rangle = \frac{1}{\sqrt{\cos^2\alpha + \sin^2\alpha \cos^2 \theta}} ( 0,\sin\alpha \cos\theta, \cos\alpha, 0)^T, \nonumber 
\end{eqnarray}
and 
\begin{eqnarray}
& &|A'_1\rangle = (0, 0, 0, 1)^T, ~~|A'_2\rangle = (0, \cos\beta, \sin\beta, 0)^T, \nonumber \\
& &|A'_3\rangle = ( 0, \sin\gamma \sin\beta, -\sin\gamma \cos\beta, \cos\gamma)^T,  \nonumber \\
& &|A'_4\rangle = (0,\sin\delta \sin\epsilon, -\sin\delta \cos\epsilon, \cos\delta)^T,  \nonumber \\
& &|A'_5\rangle = \frac{1}{\sin\delta} (0, \sin\delta \cos\epsilon, \sin\delta \sin\epsilon, 0)^T,\nonumber 
\end{eqnarray}
where we impose the conditions $\sin(\beta - \epsilon) \neq 0$, $\cos(\beta - \epsilon) \neq 0$ and $\tan\delta \tan\gamma \cos(\beta - \epsilon) = -1$.
This is analogous to the situation where one requires three qubits (dimension $8$) which is the minimal system in order to have monogamy of Bell inequality violations. Since quantum theory obeys the principle of no-disturbance, the monogamy inequality (\ref{e2}) is guaranteed to hold for these projectors. However, quantum mechanics incorporates other properties as well such as the complementarity principle \cite{Kurzynski}. For the monogamy relations of Bell inequalities, the exact trade-offs between multiple inequalities within quantum theory have been derived \cite{Kurzynski, Verstraete} using the principle of complementarity. An important open problem is to derive the exact trade-offs for contextual inequalities within quantum theory.

{\it Discussion and Conclusions.} 
Violation of a single contextual inequality (\ref{e1}) for a three-level system has been experimentally tested using a single photon \cite{Zeilinger}. The monogamy of contextuality presented here can be realized for a four-level system with feasible modifications to the existing experimental setup, and using projectors according to the commutation graphs presented in the figures above. This would establish the monogamy of contextuality as a distinct phenomenon from the monogamy of entanglement, since the notion of entanglement is not clearly applicable to a single quantum system.  

In this paper we have demonstrated that the property of monogamy so far seen for quantum correlations in composite entangled systems also carries over to single quantum systems. In particular we have shown that one can construct monogamy relations for contextual inequalities of the KCBS-type using the principle of no-disturbance. Violation of these monogamy relations is seen to be related to a violation of causality analogous to the necessity of signaling between spatially separated systems for the violation of Bell monogamy relations. The fact that Bell inequalities and contextual inequalities arise from the same origin, namely the assumption of realism, leads one to believe that features seen in the Bell scenario should carry over to the contextual scenario as well. In this regard, it would be interesting to investigate how other features such as distillation, activation of non-locality and no-signaling boxes (or Popescu-Rohrlich boxes) carry over to the contextual scenario \cite{Cabello}. 

{\em Acknowledgements}.  This research is supported by the National Research Foundation and Ministry of Education in Singapore. We acknowledge useful discussions with Tomasz Paterek and Daniel Oi.

{\bf Supplemental Material}

Here we establish Propositions $1$ and $2$ which were stated in the main text. We also illustrate the analogous relationship between the monogamy of contextuality presented and the monogamy of Bell inequalities (which are special instances of contextual inequalities) by using the commutation graph technique to derive monogamies of Bell inequality violations in no-signaling theories. 

We first introduce some graph theoretic terminology. As stated before, the independence number $\alpha(G)$ of a graph $G$ is the size of the largest independent set of $G$, where an independent set is a set of vertices of which no pair is adjacent. The vertex clique cover number $\bar{\chi}(G)$ is the minimal number of cliques needed to cover all the vertices of the graph. The clique number $\omega(G)$ is the size of the largest clique contained in the graph as a subgraph. The chromatic number $\chi(G)$ is the minimum number of colors needed to properly color all the vertices of $G$ where no two adjacent vertices have the same color. The complement of a graph $G$ is a graph $\bar{G}$ on the same set of vertices such that any two vertices in $\bar{G}$ are adjacent if and only if they are non-adjacent in $G$. A chordal graph is a graph with no induced cycles of length greater than $3$, where an induced cycle is a cycle that is an induced subgraph, i.e., between any two vertices of the subgraph there is an edge if and only if there is an edge in the original graph. A perfect graph is a graph $G$ which has the property that $\omega(G') = \chi(G')$ for all induced subgraphs $G'$ of $G$. We will show in Proposition $1$ that all chordal graphs admit a joint probability distribution, it is left open if all perfect graphs do so as well.

{\bf I. Proposition $1$:} 

A commutation graph $G$ representing a set of $n$ measurements (for any $n$) admits a joint probability distribution for these measurements if it is a chordal graph.

{\bf Proof:} 

The proof is by construction. By assumption, the commutation graph $G$ does not contain any induced cycles of length greater than $3$. Let us denote the set of vertices of $G$ by $V(G) = \{V_1, ..., V_n\}$. Let $K_3 = \{K_3^{(i)}\}$ denote the set of cycles of length $3$ or more in $G$, $K_2 = \{K_2^{(i)}\}$ denote the set of edges (of length $2$) in $G$ that are not subgraphs of any graph $K_3^{(i)}$ and $K_1 = \{K_1^{(i)}\}$ denote the set of vertices (of length $1$) in $G$ that are not subgraphs of any graph $K_3^{(i)}$ or $K_2^{(i)}$. All the edges of the chordal graph $G$ belong to one and only one of the sets $K_1$, $K_2$ or $K_3$ but they can occur more than once within a set. The vertices of $G$ may appear in more than one set and also occur multiply within each set. Let $K = K_1 \bigcup K_2 \bigcup K_3$. 

Construct the joint probability distribution for the set of $n$ measurements in $G$ as 
\begin{equation}
P(V_1,...,V_n) = \frac{\Pi_{i=1}^{\#(K_3)} \Pi_{j=1}^{\#(K_2)} \Pi_{k=1}^{\#(K_1)} P(K_3^{(i)})  P(K_2^{(j)})  P(K_1^{(k)})}{\Pi_{i < j = 1}^{\#(K)} P(K^{(i)} \bigcap K^{(j)})} 
\label{jpd}
\end{equation}

Here $\#(K)$ denotes the number of elements in set $K$ and $P(K^{(i)} \bigcap K^{(j)})$ denotes the probability of the set of vertices that are at the intersection of the two elements $K^{(i)}$ and $K^{(j)}$. We can derive as the marginal probability of this joint probability distribution, any probability $P(K^{(i)})$ (of all experimentally measurable marginals) by summing over all elements other than $K^{(i)}$ in the following manner. We first carry out the summation of all elements $K^{(j)}$ whose intersection with $K^{(i)}$ is the null set. One can immediately see that in the resulting expression all the terms in the denominator $\Pi_{i < j = 1}^{\#(K)} P(K^{(i)} \bigcap K^{(j)})$ precisely cancel with all the terms in the numerator except $P(K^{(i)})$ which completes the proof.

We now establish Proposition $2$ which is a necessary and sufficient condition for a commutation graph to yield a contextual monogamy relation. 

{\bf II. Proposition 2:} 

Consider a commutation graph $G$ representing a set of $n$ KCBS-type contextual inequalities $I_j \leq R$ ($j =1,...,n$) where each contextual inequality has non-contextual bound $R$. Then this graph gives rise to a monogamy relation using the outlined method if and only if its vertex clique cover number is $n*R$, i.e., $\bar{\chi}(G) = n*R$. 

{\bf Proof:} 

The condition that the vertex clique cover number is $n*R$ is clearly sufficient as each clique has an independence number of $1$, and cliques are the only graphs with independence number $1$. Thus, a vertex decomposition of the commutation graph into $n$ cliques gives rise to the monogamy relation, $\sum_{j = 1}^{n} I_j \leq n*R$. 

We now show that this condition is also necessary for a commutation graph to result in a monogamy relation by our method. The method relies on the vertex decomposition of the commutation graph into $m \leq n*R$ chordal subgraphs ($G_1,...,G_m$) each of which admits joint probability distribution, such that the sum of their independence numbers is $n*R$, i.e., $\sum_{j = 1}^{m} \alpha(G_i) = n*R$. Now, all chordal graphs are known to be perfect, i.e., the size of the largest clique in every induced subgraph of the chordal graph equals the number of colors needed to color that induced subgraph \cite{Berge}. For all perfect graphs $G$, $\alpha(G)$ can be shown to be equal to $\bar{\chi}(G)$, by the following series of equalities. Note that from the definition of a perfect graph $\omega(G) = \chi(G)$ by taking the induced subgraph of $G$ to be $G$ itself. That is, the size of the largest clique in the graph is the minimum number of colors needed to properly color $G$. For all perfect graphs $G$, their complement graph is also perfect by the Weak Perfect Graph Theorem \cite{Lovasz}, i.e., $\omega(\bar{G}) = \chi(\bar{G})$. It is readily seen that for any graph $G$, $\omega(\bar{G}) = \alpha(G)$, i.e., the independence number of a graph is equal to the size of the largest clique in its complement graph. Also, $\chi(\bar{G}) = \bar{\chi}(G)$ meaning that the number of colors required to color a graph $G$ is equal to the number of cliques that cover $\bar{G}$. From the above equalities, the following result is obtained that for all perfect graphs, and in particular for chordal graphs $G$, 
\begin{equation}
\alpha(G) = \bar{\chi}(G),
\label{berge1}
\end{equation}
the independence number of a chordal graph is equal to its vertex clique cover number. 

Now, by eqn.(\ref{berge1}), in the derivation of the contextual monogamy relation by vertex decomposition into chordal subgraphs, $\sum_{j = 1}^{m} \alpha(G_i) = \sum_{j=1}^{m} \bar{\chi}(G_i) = n*R$. That is, the chordal subgraphs can be vertex decomposed further exactly into a set of $n*R$ cliques. This proves that the condition that the vertex clique cover number be equal to $n*R$ is both necessary and sufficient for a commutation graph to result in a contextual monogamy relation by the method outlined in the text. As explained in the text, one can readily check whether a given graph gives rise to a contextual monogamy relation by using an algorithm to find its vertex clique cover number and checking if it is equal to $n*R$. If we have a set of $n$ contextual inequalities such that $n_1$ of them have a bound of $R_1$, $n_2$ of them have a bound of $R_2$, and in general $n_k$ of them have a bound of $R_k$, then it is easy to see that the proof extends to give the necessary and sufficient condition as the vertex clique cover number being equal to $\sum_k n_k*R_k$, i.e., $\bar{\chi}(G) = \sum_k n_k*R_k$. Note that we are not interested in the case that $\bar{\chi}(G) < n*R$ since as explained in the text, this gives rise to monogamies even within non-contextual theories. The application of Proposition $2$ to the graphs in the main text gives that each of the graphs there have vertex clique cover number of $4$ leading to the monogamy relations. The above Proposition gives the condition for monogamies for KCBS-type contextual inequalities (of the form $\sum_i p(A_i = 1) \leq R$ with $R$ being the independence number of the associated graph). An interesting open question is to identify the conditions under which other types of contextual inequalities give rise to monogamy relations as well.

{\bf III. Bell monogamy relations from commutation graphs:}

Finally, we show the analogous nature of contextual and Bell monogamy relations by using the commutation graph techniques to show the Bell monogamy relation for the simplest case of three observers, Alice, Bob and Charlie. Here Alice and Bob try to violate a Bell-CHSH inequality while Alice and Charlie simultaneously try to violate the same Bell-CHSH inequality with Alice using the same settings in both cases. 

\begin{figure}[h]
\begin{center}
\includegraphics[scale=0.3]{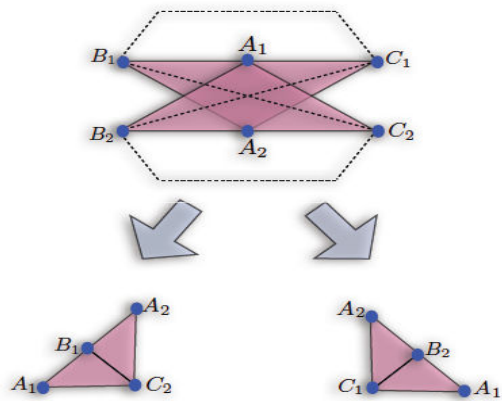}
\end{center}
\vspace{-1cm}
\caption{The commutation graph (top) and its decomposition (bottom) for the Bell-CHSH monogamy relation for Alice, Bob and Charlie.}
\label{figBell}
\end{figure}

Similar derivations can be carried out for Bell inequalities involving more measurement settings as well. Consider the violation of a Bell-CHSH inequality involving two measurement settings for each of two spatially separated systems labeled Alice and Bob, and also for the two separated systems Alice and Charlie. Denoting the measurements performed by Alice as $A_1$ and $A_2$ (those by Bob as $B_1$ and $B_2$ and by Charlie as $C_1$ and $C_2$), the spatial separation guarantees that any set of measurements $A_i$, $B_j$, $C_k$ are compatible (though not mutually exclusive) and that measurement pairs $\{A_1, A_2\}$, $\{B_1, B_2\}$ and $\{C_1, C_2 \}$ cannot be jointly performed. The commutation graph depicting this scenario is presented in Fig. \ref{figBell} where edges denote joint measurability, though not exclusiveness. The vertices of the graph denote the measurements whose outcomes take values between $+1$ and $-1$ unlike the $0/1$ outcomes considered in the contextuality scenario. A monogamy relation can be derived for the two Bell-CHSH inequalities with local realistic bounds $R$, $\it{\bf{B}}(A_1, A_2, B_1, B_2) = <A_1 \otimes B_1> + <A_1 \otimes B_2> + <A_2 \otimes B_1> - <A_2 \otimes B_2> \leq R$ and $\it{\bf{B}}(A_1, A_2, C_1, C_2) \leq R$  from this graph as follows. 

We first note that this graph can be decomposed into two sub-graphs of four vertices, each of which represents a single Bell inequality, namely $A_1, A_2, B_1, C_2$ and $A_1, A_2, B_2, C_1$ where we ignore edges in the original graph connecting the two resulting sub-graphs. A joint probability distribution reproducing all measurable marginals can be constructed for each of these sub-graphs, for instance for the sub-graph $A_1, A_2, B_1, C_2$ we can construct $p(a_1, a_2, b_1, c_2) = p(a_1, b_1, c_2) p(a_2, b_1, c_2)/p(b_1, c_2)$, where the terms on the right-hand side are guaranteed to exist due to the joint measurability. Notice that the no-signaling principle is crucial to the above derivation as it ensures that $p(b_1, c_2)$ derived as the marginal probability from $p(a_1, b_1, c_2)$ is the same as that from $p(a_2, b_1, c_2)$. Therefore, each of the Bell inequalities represented by $A_1, A_2, B_1, C_2$ and $A_1, A_2, B_2, C_1$ cannot be violated in any theory obeying the no-signaling principle. Hence, a monogamy relation holds for the Bell inequality violations, $\it{\bf{B}}(A_1, A_2, B_1, B_2) + \it{\bf{B}}(A_1, A_2, C_1, C_2) \equiv \it{\bf{B}}(A_1, A_2, B_1, C_2) + \it{\bf{B}}(A_1, A_2, C_1, B_2) \leq 2 R$ \cite{Pawlowski}. The construction can be readily extended to more systems and more measurements as well.




\end{document}